# Simulation of corrosion and mechanical degradation of additively manufactured Mg scaffolds in simulated body fluid

Mohammad Marvi-Mashhadi[1], Wahaaj Ali[1,2], Muzi Li[2], Carlos González[2,3], Javier LLorca[2,3]

[1]*Carlos III University of Madrid, Av. de la Universidad 30, 28911 Leganés, Madrid, Spain*

[2]*IMDEA Materials Institute, C/Eric Kandel 2, 28906 Getafe, Madrid, Spain,*

[3]*Department of Materials Science, Polytechnic University of Madrid, 28040 Madrid, Spain*

**Abstract**

A simulation strategy based in the finite element model was developed to model the corrosion and mechanical properties of biodegradable Mg scaffolds manufactured by laser power bed fusion after immersion in simulated body fluid. Corrosion was simulated through a phenomenological, diffusion-based model which can take into account pitting. The elements in which the concentration of Mg was below a certain threshold (representative of the formation of $Mg(OH)_2$) after the corrosion simulation were deleted for the mechanical simulations, in which Mg was assumed to behave as an isotropic, elastic-perfectly plastic solid and fracture was introduced through a ductile failure model. The parameters of the models were obtained from previous experimental results and the numerical predictions of the strength and fracture mechanisms of WE43 Mg alloy porous scaffolds in the as-printed condition and after immersion in simulated body fluid were in good agreement with the experimental results. Thus, the simulation strategy is able to assess the effect of corrosion on the mechanical behavior of biodegradable scaffolds, which is critical for design of biodegradable scaffolds for biomedical applications.

# 1. Introduction

Current state-of-the-art implants for bone healing are fabricated by additive manufacturing from biocompatible Ti and Co alloys or stainless steels [1-3]. These metallic alloys are biocompatible and provide the necessary load bearing capability while additive manufacturing allows the fabrication of patient-specific scaffolds and implants from computer-assisted tomography medical images in orthopaedics and traumatology [4]. However, they are not biodegradable and, thus, a second surgery is often needed to remove the implants once their purpose is fulfilled because they might lead to permanent physical irritation/or and chronic inflammatory local reactions. These limitations can be overcome through the use of biodegradable metals (Fe, Mg, Zn) that allow the progressive degradation of the scaffold *in vivo* while the corrosion products are eliminated and/or metabolised. As a result, the implant dissolves completely after tissue healing with no implant residues [5-6].

Fabrication of Mg scaffolds with controlled topology and pore structure (size, shape and connectivity) was recently demonstrated by different groups by laser powder bed fusion (LPBF) [7-12]. This strategy opens the possibility for a new generation of bone implants because the biocompatibility and osteopromotive properties of Mg can stimulate new bone formation [13]. However, one key issue to ensure the successful clinical application of porous biodegradable implants is to tune the degradation rate. Very high degradation rates may lead to the massive release of degradation products (hydrogen in the case of Mg alloys) that might trigger unwanted reactions with surrounding tissue (inflammation or necrosis) [14-16] and even premature failure of the implants [12]. This problem is more important in the case of porous scaffolds because of the complex shape and of the large surface area. In addition, biodegradable implants should be able to fulfil initially the mechanical function to allow implant stabilization during the inflammation and tissue repair stages. Afterwards, they should slowly degrade as bone tissue heals and replaces the implant gradually over time (> 6 months [17-18]). Ideally, the stiffness of the bone-implant compound should remain constant during the healing period as the load gradually transfers from implant to the bone [19]. In this respect, coatings can be used to tailor the corrosion rate and also to enhance cell affinity and growth, which in turn results in better bone-scaffold integration [20].

The optimization of scaffold degradation by means of experimental trial-and-error approaches is very expensive and time-consuming and, thus, numerical simulations are very much in need to guide scaffold design [21]. Nevertheless, Mg corrosion under loading in a physiological environment is an extremely complex problem that involves different mechanisms (micro-galvanic reactions, uniform and localized corrosion, stress corrosion cracking, corrosion fatigue [22]). As a result, corrosion rates depend on the alloy composition and heat treatment, stress level as well as on the features of physiological fluid and, thus, phenomenological models are preferred to simulate corrosion and its effect on the mechanical behavior. For instance, Gastaldi et al. [23] proposed a simple approach to analyze the effect of corrosion on the mechanical properties of Mg stents. It was assumed that surface corrosion was due to two contributions, one uniform associated with the electrochemical effects and another depending on the local stress. Both contributions were added to a damage parameter that reduced the mechanical properties of the corroded zone following the principles of continuum damage mechanics. The approach was modified by Grogan et al. [24] to take into account the effect of pitting corrosion on the degradation of Mg stents. To this end, the corrosion rate at each point of the surface was dependent on a random number obtained from a Weibull distribution. Later, Grogan et al. [25] simulated corrosion in Mg stents using a physically-based model in which the corrosion rate was governed by the diffusion of Mg ions in solution. The same physical model was used by Gartzke et al. [26] to simulate diffusion-controlled corrosion of porous scaffolds using the level set method to track corrosion. The corrosion results were used as input of a simple mechanical model in which the elastic modulus in each point of the scaffold decreased proportionally to the concentration of Mg. Thus, they were able to provide estimations of the degradation of the scaffold stiffness with corrosion time.

To the authors' knowledge, simulations of corrosion in Mg scaffolds with complex geometries are limited to very simple models that assume uniform corrosion controlled by diffusion [7]. Nevertheless, several recent experimental studies of corrosion in simulated body fluid (SBF) of 3D Mg scaffolds manufactured by LPBF have demonstrated that pitting corrosion is very important [9, 11-12]. Moreover, the overall corrosion rate as well as the degree of corrosion localization depends on the processing route [9], heat treatment [12] and surface modifications [11]. Thus, the design of porous Mg scaffolds manufactured by LPBF requires a simulation tool that is able to handle these different scenarios and determine the effect of corrosion on the mechanical properties as a function of time and this is the objective of this investigation. To this end, corrosion is simulated using a diffusion-controlled model that can be applied to any three-dimensional geometry, following the methodology developed by Gartzke et al. [26] and incorporates the approach developed by Grogan et al. [24] to account for pitting corrosion. The parameters of the corrosion model were calibrated against experimental results obtained by X-ray computed tomography (XCT) in WE43 Mg scaffolds manufactured by LPBF immersed in SBF [12]. Afterwards, the mechanical properties of the as-printed scaffolds were computed using the finite element model of the corroded scaffolds. The Mg alloy was assumed to follow an elasto-plastic behavior and failure was dictated by a ductile failure criterion and the fully-corroded regions of the scaffolds were not included in the simulations. The mechanical properties of the scaffolds in compression obtained from the simulations were compared with experimental results from a previous investigation to assess the validity of the approach.

## 2. Material and experimental techniques

The details of the scaffold fabrication, corrosion and mechanical tests can be found in [12] and only the most relevant details are presented here for the sake of completion. Cubic scaffolds of WE43 Mg alloy of nominal dimensions of 10 × 10 × 10 mm$^3$ based on a unit cell with a body-centred cubic structure were fabricated by LPBF at Meotec GmbH (Aachen, Germany). Scaffolds with porosities of 63% and 58% and strut diameters of 518 ± 27 μm and 793 ± 39 μm, respectively, were fabricated. Corrosion tests were carried out in the scaffolds with struts of 518 μm in diameter by immersion in Dulbecco's Modified Eagle Medium SBF. The degradation rate of the as-printed scaffolds was measured from XCT in a Phoenix Nanotom after 3, 5 and 7 days of immersion. The sample-focus distance was set to 25 mm and the detector-focus distance to 200 mm, leading to a resolution of 6.25 μm per pixel. The corrosion products and the Mg alloy have different X-ray absorption and the total volume of Mg remaining in the scaffolds could be easily determined by binarizing the information in the 3D tomograms with a proper threshold using ImageJ.

Compression tests of the as-printed and corroded scaffolds were carried out with a Kammrath & Weiss GmbH micromechanical testing machine at a crosshead speed of 0.6 mm/min up to a maximum engineering strain of 60%. The nominal stress $S$ was determined from the load divided by the initial cross-section of the scaffolds while the engineering strain ε was calculated as the crosshead displacement of the mechanical testing machine divided by the initial height of the sample.

## 3. Corrosion model

Corrosion was simulated through a diffusion-based model able to capture mass loss by pitting corrosion on three-dimensional biodegradable Mg scaffolds. The diffusion process in the scaffold is simulated through the finite element method with Abaqus/explicit [27] according to Fick's law

$$\frac{\partial c}{\partial t} = \nabla \cdot (D\nabla c) \qquad (1)$$

where $c$ is the concentration of Mg, $t$ the time, $D$ the isotropic diffusion coefficient of Mg in the alloy ($10^{-8}$ mm$^2$/s) and $\nabla$ the gradient function. The scaffold is discretized with 4-node linear tetrahedra (DC3D4) with an average length of 0.1 mm and it is assumed that corrosion begins at each node on the surface of the scaffold. To this end, a boundary condition of concentrated outward point flux is applied to each node $i$ on the scaffold surface (Fig. 1)-where the flux at node $i$ is given by

$$J_i = \beta f(x_i) = \beta \gamma (x_i)^{\gamma-1} e^{-(x_i)^\gamma} \qquad (2)$$

where $f(x_i)$ is the Weibull probability density function and $x_i$ is a random number.

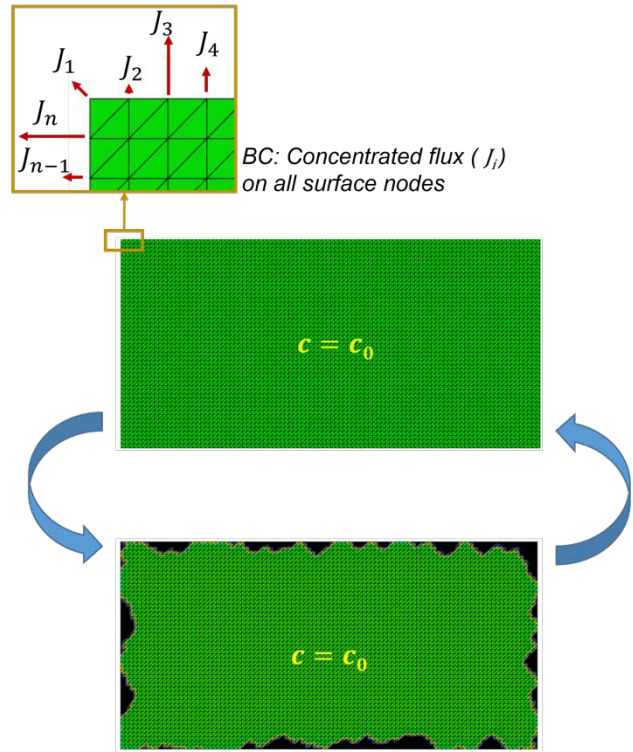

Figure 1. Schematic of the simulation of the corrosion process (including pitting) through the finite element simulation of the diffusion of Mg. The initial concentration of Mg in the solid is $c_0$ and a random flux of Mg is applied to boundaries following a Weibull probability density function in each time step. The equilibrium concentration of Mg in the solid is determined from the finite element solution of eq. (1) and the process is repeated until the desired time is reached.

The flux of Mg in each node of the discretization is distributed among the finite elements in the scaffold surface that concur to the node proportionally to the element surface area. In this phenomenological model, the parameter $\beta$ defines the average corrosion rate and $\gamma$ controls the degree of localization of corrosion, i.e. the lower $\gamma$ the higher pit formation [24]. Thus, $\beta$ and $\gamma$ have to be calibrated for each alloy and environment to account for the electrochemical and microstructural factors (pH, alloy composition, precipitates, etc.) that determine the corrosion of the alloy.

The initial concentration ($c_0$) of Mg in all the elements of the scaffold at the beginning of the simulation is $6.487 \times 10^{-5}$ mol/mm$^3$, which is equal to the concentration of Mg in the WE43 alloy. Diffusion leads to a reduction in the Mg concentration in each element but the diffusion simulations assume that corroded elements are attached to the scaffold. Thus, the point flux is applied to the nodes on the initial scaffold surface and no new surfaces are created due to element removal during the diffusion analysis. Nevertheless, the elements whose Mg content at the end of the diffusion simulation is $\leq 0.4c_0$ are assumed to be fully corroded because the Mg content of the final corrosion product Mg(OH)$_2$ is 40%. Thus, only these elements were taken into account to calculate the mass loss, that was used to calibrate the parameters $\beta$ and $\gamma$ from the XCT tomograms of the corroded scaffolds for different immersion times in SBF. Moreover, these elements were deleted from the finite element mesh that was used to simulate the mechanical behavior, assuming that the mechanical properties of the corroded material can be neglected [28]. The simulation of the corrosion process in one scaffold during 168 took one hour in Intel a Xeon W3530 @ 2.8 GHz with 4 CPUs.

## 4. Mechanical model

Numerical simulations of the mechanical behavior of as-printed and corroded scaffolds were also carried out with Abaqus/Explicit [27] within the framework of the theory of finite deformations and with the initial unstressed state as reference. The scaffold was discretized with C3D4 linear tetrahedra. The WE43 Mg alloy was modelled as an isotropic, elastic-perfectly plastic solid following the J$_2$ theory of plasticity. The elastic modulus $E$ (= 44 GPa), the Poisson's ratio $\nu$ (= 0.27) and the yield stress $\sigma_y$ (= 200 MPa) of the WE43 Mg alloy were obtained from the literature [29]. Stress-strain curves for WE43 Mg alloy manufactured by LPBF are not available in the literature and there are important differences in the microstructure of wrought and LPBF WE43 Mg alloys, i.e. the presence of a large volume fraction of oxide particles in the latter, which leads to failure a low strains [30]. Thus, it was more important to include the effect of fracture in the model (which was calibrated through the information in [29] and the comparison with the experimental results, as shown below) than strain hardening. Plastic anisotropy was not considered in the model because the experimental analysis of the deformed scaffolds showed that twinning deformation was negligible [12].

Fracture was introduced in the simulations through a ductile damage model available in Abaqus [27]. The model assumes that the stress tensor in the material, $\sigma_{ij}$, is given by the scalar damage equation

$$\sigma_{ij} = (1-d)\bar{\sigma}_{ij} \qquad (3)$$

where $d$ is the overall damage variable and $\bar{\sigma}_{ij}$ the effective (or undamaged) stress tensor computed in the current increment. In addition, the damage model also assumes that the elastic modulus is proportional to $1-d$. Thus, the material has lost its load-carrying capacity when $d = 1$ and the finite elements in which this condition is fulfilled are removed from the mesh.

Damage initiation is controlled by an internal variable, $\omega_D$, which increases monotonically with plastic deformation according to

$$\omega_d = \int \frac{d\bar{\varepsilon}^{pl}}{\bar{\varepsilon}_d^{pl}(\eta)} \qquad (4)$$

where $\bar{\varepsilon}^{pl}$ stands for equivalent plastic strain and $\bar{\varepsilon}_d^{pl}$ is a model parameter that indicates the critical equivalent plastic strain at the onset of damage which is a function of stress triaxiality, η. Damage begins when $\omega_d = 1$. The dependence of $\bar{\varepsilon}_D^{pl}$ with the stress triaxiality under quasi-static loading conditions was measured by Kondori and Benzerga [29] for a wrought WE43 Mg alloy and it is shown in Table 1.

The dependence of $\bar{\varepsilon}_D^{pl}$ with the stress triaxiality under quasi-static loading conditions was measured by Kondori and Benzerga [29] for a wrought WE43 Mg alloy and it is shown in Table 1. The triaxiality was defined as the ratio between the hydrostatic and the deviatoric stresses. The mechanical response of the calibrated model for the WE43 Mg alloy in tension showed a constant flow stress of 200 MPa up to 6% strain, when damage began. Wrought WE43 alloys show higher ductility ($\approx$ 13% [29]) but it is reasonable that the tensile ductility of the WE43 Mg alloy processed by LPFB is only 6% because of the large volume fraction of oxides and porosity associated with the processing technique.

Table 1. Dependence of $\bar{\varepsilon}_D^{pl}$ with the stress triaxiality for a wrought WE43 Mg alloy [29].

| Triaxiality | 0 | 0.33 | 0.66 | 1 | 1.33 |
|---|---|---|---|---|---|
| $\bar{\varepsilon}_D^{pl}$ | 0.15 | 0.13 | 0.15 | 0.11 | 0.05 |

The evolution of the damage variable $d$ after the onset of damage is calculated as

$$\dot{d} = \frac{L\dot{\bar{\varepsilon}}^{pl}}{\bar{u}_d^{pl}} \tag{5}$$

where $L$ is the characteristic element length and $\bar{u}_d^{pl}$ the critical equivalent plastic displacement, another parameter of the model that controls the energy dissipated until fracture. $L$ is equal to the element size in the case of first order elements and it is introduced in the formulation of the damage model to reduce the influence of the element size on the mechanical response. The magnitude of $\bar{u}_d^{pl}$ = 0.03 mm in the model was chosen to obtain good agreement between the experiments and simulations of the as-printed scaffolds and is associated with the fracture energy of the Mg alloy, $G_F$, that can be estimated from a simple tensile test until failure according to [27]

$$G_F = \int_{\bar{\varepsilon}_0^{pl}}^{\bar{\varepsilon}_f^{pl}} L\, \sigma_y\, d\, \bar{\varepsilon}^{pl} = \int_0^{\bar{u}_d^{pl}} \sigma_y\, d\, \bar{u}^{pl} \tag{6}$$

where $\sigma_y$ is the applied tensile stress. This simulation leads to $G_F$ = 8.4 kJ/m², which is a reasonable value for a Mg alloy. The simulations of the mechanical behavior until failure of the as-printed scaffolds with nominal strut diameter of 500 µm and 750 µm took 15 and 21 hours, respectively, in an Intel Xeon Gold 6126 @ 2.6 GHz with 20 CPUs.

**5. Simulation results and comparison with experiments**

The STP files used as input to manufacture the scaffolds were also used to generate the finite element models for the corrosion and mechanical simulations using Abaqus/CAE. The strut diameter in the STP files were 500 µm and 750 µm, very close to the experimental ones. The same discretization with linear tetrahedra (4 nodes) was used for both types of simulations.

*5.1 Corrosion*

Diffusion simulations were carried out in the scaffold with a nominal strut diameter of 500 μm. An initial set of simulations was carried out to assess the relationship between the model parameters $\beta$ and $\gamma$ and the mass loss $\Delta m$ of the scaffold after 168 hours of immersion in SBF and the results are plotted in Fig. 2. As indicated above, the mass loss is determined from elements in which the concentration of Mg is below 40% of the initial concentration. Obviously, $\Delta m$ increases with $\beta$ for a given value of $\gamma$ as $\beta$ dictates the average flux of Mg for a given concentration gradient. Nevertheless, the influence of $\beta$ in the corrosion rate is modulated by $\gamma$, as it is shown in Fig. 2: the higher $\gamma$, the higher variation in the overall mass loss in the scaffold with $\beta$.

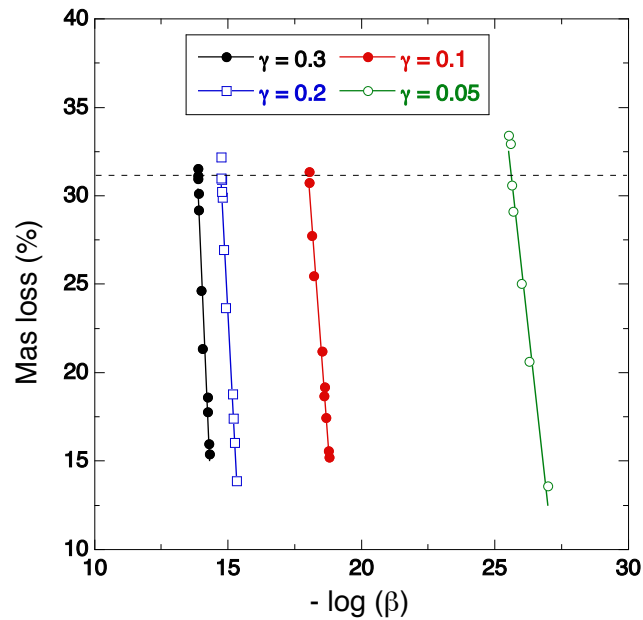

Figure 2. Simulation results of the mass loss of the scaffold (expressed as % of the initial mass) after 168 hours of immersion in SBF as a function of the parameters $\beta$ and $\gamma$ of the diffusion model. $\beta$ is expressed as mol s$^{-1}$ mm$^{-2}$. The horizontal dashed lines indicated the mass loss measured by XCT.

Four pairs of $\beta$ and $\gamma$ were chosen to simulate the corrosion of the scaffold under the assumption that the total mass loss predicted by the corrosion model was equal to the one measured by XCT after 168 hours (Fig. 3). The four pairs of parameters were able to predict with reasonable accuracy the evolution of the mass loss with time for the scaffold but they led to diverse corrosion patterns as a result of the differences in the Weibull parameter $\gamma$ that controls that degree of pitting. The progress of corrosion with time in the scaffold is depicted in Fig. 4 for the simulations with $\beta = 1.29 \times 10^{-14}$ mol s$^{-1}$ mm$^{-2}$ and $\gamma = 0.3$. In this figure, the elements in which the concentration of Mg was below $0.4c_0$ have been deleted to show the progress of corrosion. They show that corrosion is not uniform but begins by pitting corrosion on localized points of the scaffold surface. After 168 hours, corrosion has spread over the whole scaffold and some struts have been fully corroded, compromising the structural stability of the scaffold.

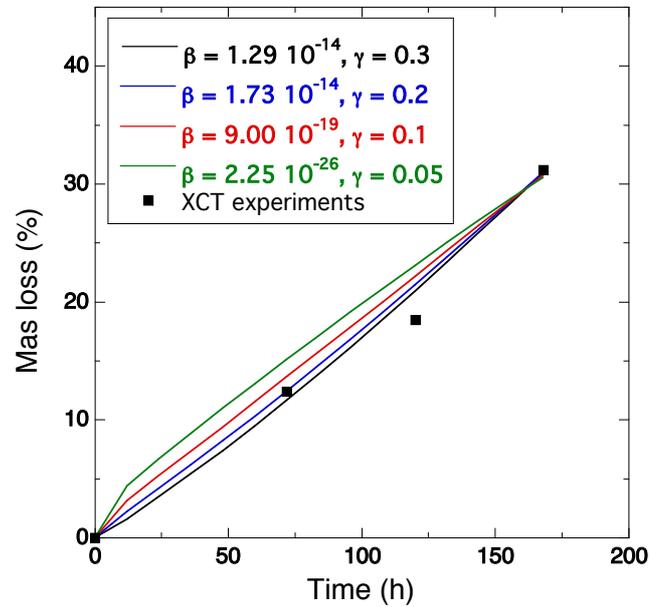

Figure 3. Mass loss (expressed as % of the initial mass of the scaffold) vs. time (h) for the scaffold with struts of 500 μm nominal diameter immersed in SBF. Solid squares stand for the experimental results obtained by XCT [12]. The lines represent the predictions of the diffusion model for different pairs of $\beta$ and $\gamma$. $\beta$ is expressed as mol s$^{-1}$ mm$^{-2}$.

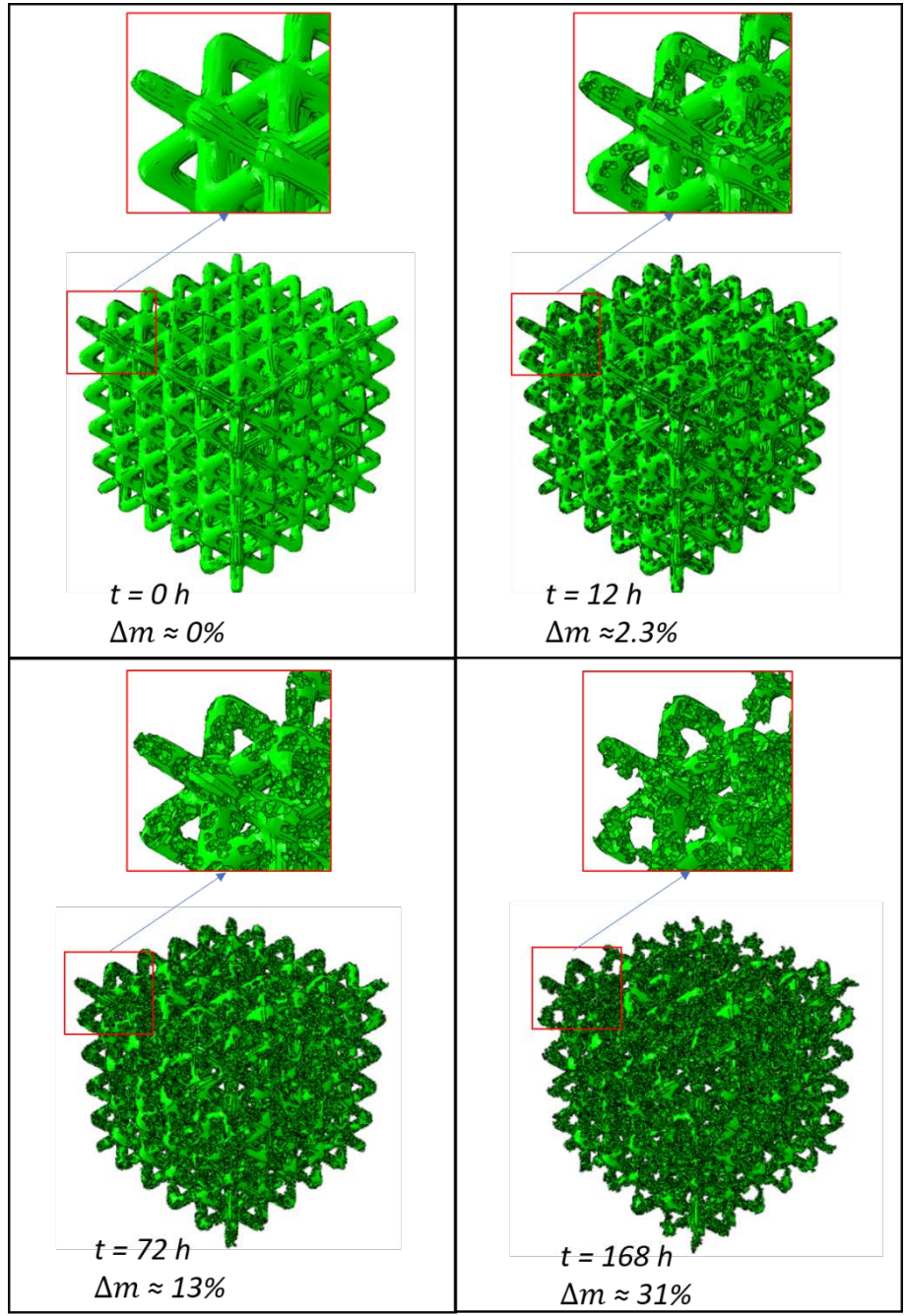

Figure 4. Simulations of corrosion in the scaffold with 500 μm nominal strut diameter as a function of the immersion time in SBF. The elements in which the concentration of Mg was below $0.4c_0$ have been deleted. The immersion time $t$ (in hours) and the mass loss $\Delta m$ (%) are indicated below each plot. $\beta = 1.29 \times 10^{-14}$ mol s$^{-1}$ mm$^{-2}$ and $\gamma = 0.3$.

The influence of the Weibull parameter on the corrosion process in three different section of the scaffold (1, 2 and 3) is shown in Fig. 5. The black regions correspond to fully corroded parts of the scaffold in which $c \geq 0.4c_0$ while the remaining zones are green. The mass loss was 31% in all cases but localization of corrosion was controlled by $\gamma$. Corrosion was fairly uniform when $\gamma = 0.3$ whereas large differences in the corrosion rate among different zones of the scaffold can be seen when $\gamma = 0.05$.

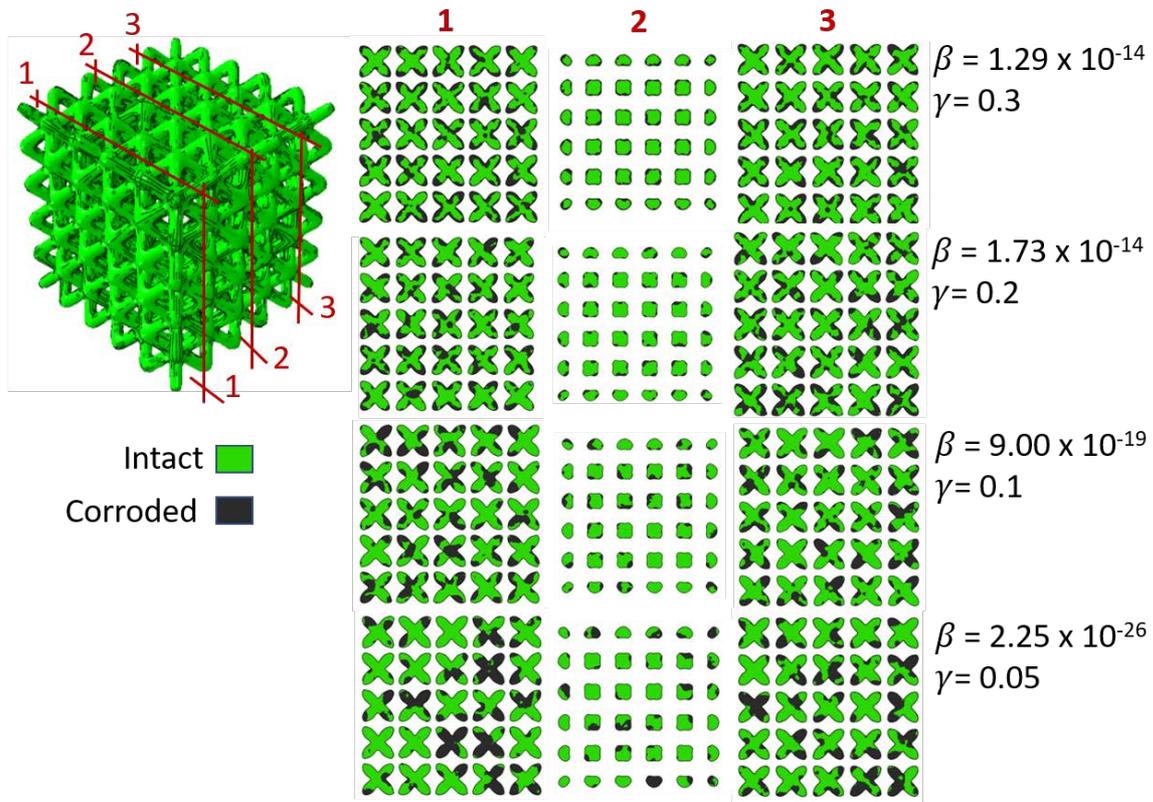

Figure 5. Simulated cross-sections of the scaffold with 500 μm nominal strut diameter after 168 hours of immersion in SBF for different values of $\beta$ (expressed as mol s$^{-1}$ mm$^{-2}$) and $\gamma$. The elements in which the concentration of Mg was below $0.4c_0$ have been deleted. The total mass loss was 31% in all cases.

The as-printed WE43 Mg scaffolds with 500 μm nominal strut diameter were analyzed by means of XCT after 3, 5 and 7 days in SBF and the videos of the corresponding XCT tomograms are shown in videos S1.mp4, S2.mp4 and S3.mp4 in the supplementary material. The white regions correspond to the uncorroded WE43 Mg alloy while the grey regions stand for Mg(OH)$_2$, that has a different adsorption of the X-rays. The corrosion of Mg to Mg(OH)$_2$ leads to a large increase in volume ($\approx$ 78% [31]). Due to the corrosion, some regions of the scaffold have fallen off and, thus, do not appear in the tomogram. In general, corrosion of the scaffolds was highly localized, particularly in the central region of the scaffold and the experimental data were better reproduced by the corrosion simulations with $\gamma = 0.05$, that were chosen as input for the mechanical simulations.

The exact reason for this localization of corrosion at the center of the scaffold is unclear but several hypotheses can be advanced. The analysis of the as-printed scaffolfds by XCT revealed the presence of a larger volume fraction of unmelted Mg particles at the center of the scaffold [12]. They lead to rough surfaces that can enhance the corrosion rate due to pit corrosion. In addition, the diffusion of $Mg^{++}$ ions from the center of the scaffold is hindered by the narrow channels, leading to a high $Mg^{++}$ ion concentration. These ions attract $Cl^-$ ions to balance the electrical charge but $Cl^-$ attacks the passive oxide film on the surface and increases the corrosion rate. And finally, the passive oxide layer on the periphery of the scaffold could act as cathode while the unprotected material as the center behaves as an anode, further increasing the corrosion rate at the center of the scaffold.

*5.2 Mechanical properties*

The mechanical behavior in compression of the scaffolds with struts of 500 µm and 750 µm in diameter as well as of the corroded scaffolds with struts of 500 µm in diameter was simulated using Abaqus/Explicit [27] (Fig. 6). The scaffolds were compressed between two rigid plates using the General Contact Algorithm in to avoid that the elements of the model overlap with the boundaries of cubic scaffold as a result of buckling. The velocity of the upper surface was -0.033 mm/s, which corresponds to a nominal strain rate of $0.0033\ s^{-1}$. The friction coefficient between the plates and the Mg struts was 0.6. It was assessed that the explicit analysis was carried out under quasi-static conditions: the kinetic energy was always less than 2% of the internal energy and the reaction forces extracted from reference nodes on the loading plates were symmetric throughout the analysis.

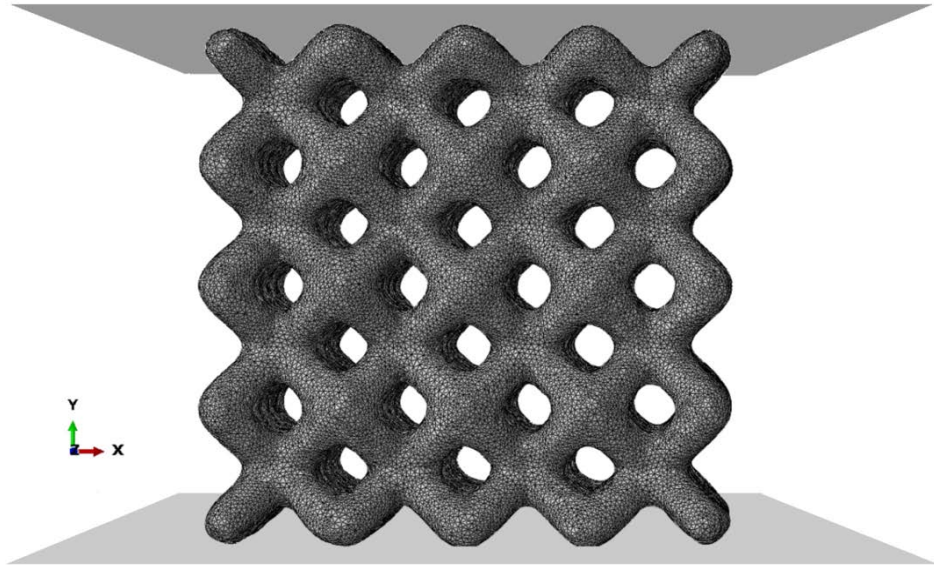

Figure 6. Schematic of the compression of the scaffold between two rigid plates.

The nominal stress $S$ was calculated as the applied load divided by the initial cross-section of the scaffold and the engineering strain $\varepsilon$ was determined from the displacement of the upper rigid plate divided by the initial height of the scaffold. The $S-\varepsilon$ curves obtained from the simulations of the as-printed scaffolds with struts of 500 μm and 750 μm are plotted in Figs. 7a and 7b, respectively, together with the experimental $S-\varepsilon$ curves from [12]. Two simulation curves are plotted in each figure. The dashed blue lines correspond to the simulations in which the Mg is assumed to behave as an elastic-perfectly plastic solid without damage. The red lines stand for the mechanical behavior using the ductile damage model for Mg. Both curves are superposed up to an applied strain of ≈10-15%. Afterwards, the simulation without damage shows continuous hardening due to the progressive densification of the scaffold while the presence of damage leads to strong oscillations in the $S-\varepsilon$ curves. The experimental $S-\varepsilon$ curves for both scaffolds are close to the simulation results with damage which reproduced very accurately the initial yield strength and the average stress carried by the scaffold during deformation. The elastic moduli of the scaffolds in the simulations were larger than the experimental ones. These differences arise from the experimental difficulties because it is practically impossible to ensure that both surfaces of the scaffolds are perfectly parallel and that the loading plates are also perfectly parallel to the surfaces. These misalignments lead to localized contacts between the plate and the scaffold during the experiment which reduces the nominal elastic modulus measured from the slope of the linear region of the $S-\varepsilon$ curve. These problems are much more dramatic in the case of porous scaffolds manufactured by LPBF because the planes determined by the scaffold nodes in both surfaces will never be parallel and -of course- lead to very large errors in the elastic modulus after corrosion, which destroys any parallelism between both surfaces and between the surfaces and the loading plate.

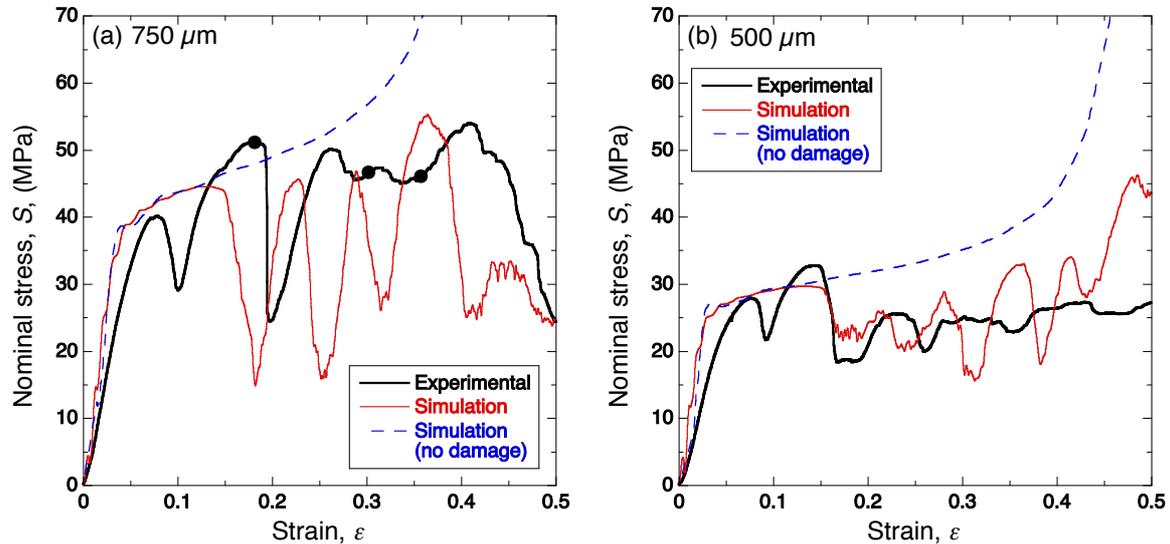

Figure 7. Nominal stress (S) vs. strain ($\varepsilon$) curves of the as-printed scaffolds deformed in compression. (a) Scaffold with strut diameter of 750 µm. (b) Scaffold with strut diameter of 500 µm.

The video of experimental test in the scaffold with strut diameter of 750 µm can be found in the supplementary material (video S4.mp4) and optical images of the scaffold extracted from the video are plotted in Fig. 8 for different values of the applied strain. They show that the first stress drop at an applied strain of 10% was associated with damage in the first row of cells by cracks parallel to the loading axis. The cracks were nucleated at the joints of four struts of the bcc lattice. Tensile stress perpendicular to the loading axis developed at the joints as a result of compressive loading, leading to vertical cracks while the struts were mainly loaded in bending. Afterwards, the stress increased, reaching a maximum of ≈ 50 MPa, and it was followed by another stress drop by the fracture of struts at the bottom row of cells when the applied strain reached 20%. Thus, the serrated shape of the $S - \varepsilon$ curve is due to progressive collapse of horizontal layers beginning by those closer to the top and bottom rigid plates.

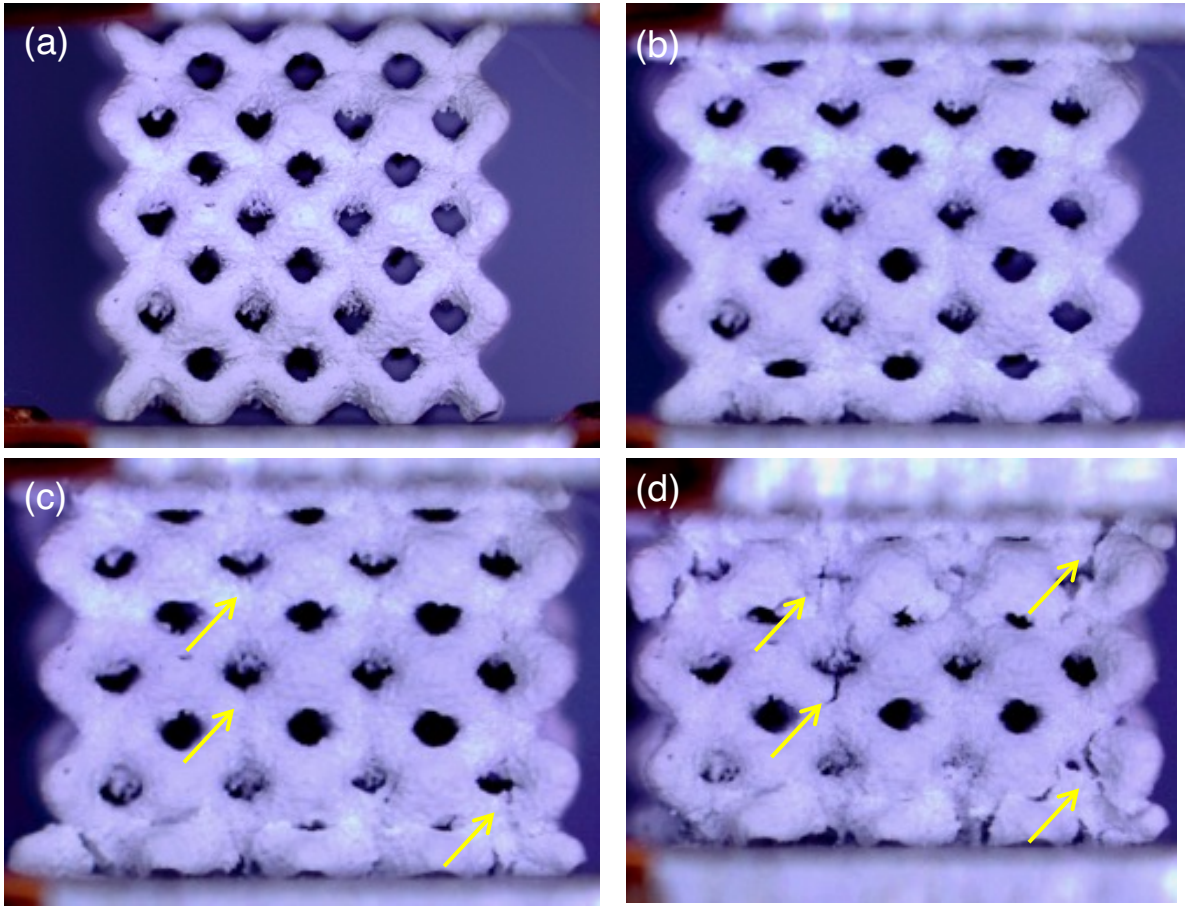

Figure 8. Optical micrographs of the scaffold with nominal strut diameter of 750 μm deformed up to different strains marked by black circles in Fig. 6a. (a) 0%. (b) 18%. (c). 30%. (d) 36%. Cracks at the joints parallel to the loading direction are indicated by the yellow arrows.

The deformation mechanisms in the simulations are equivalent and the simulated $S - \varepsilon$ curve also shows periodic serrations that are associated with the progressive failure of horizontal rows of cells in the scaffold, as shown in the snapshots of the simulations in Fig. 9. The deformation of the scaffold is uniform up to 10% strain (Fig. 9a) but damage is localized afterwards in the central row of the scaffold, as shown by the cracks (marked in red) that appear in the lateral views of the deformed scaffold (Figs. 9b and c). It should be noticed that the stress state in the simulated scaffold is not perfectly uniaxial because of the friction between the loading plates and the scaffold. Thus, compressive stresses perpendicular to the loading axis are present in the simulations. They are higher close to the loading plates and are gradually relieved with the distance to the surface and this is the reason why damage always begin in the central section in the simulations of perfect scaffolds. In addition, the lateral constraint induced by the friction with the loading surfaces leads to small differences in the stresses across any section of the scaffold. The local stresses in the centre of the scaffold are slightly higher than at the boundary and damage is triggered at the centre.

Further deformation after the first row of cells has collapsed leads to the failure of a second row of cells (the one on top of the first collapsed row) at an applied strain of 25% (Figs. 9d and e), which is associated with the second stress drop in Fig. 6a. Vertical cracks in the simulations and in the experiments appear at the joints among struts because the kinematics of compressive deformation of the bcc lattice leads to the development of horizontal tensile stresses at these points.

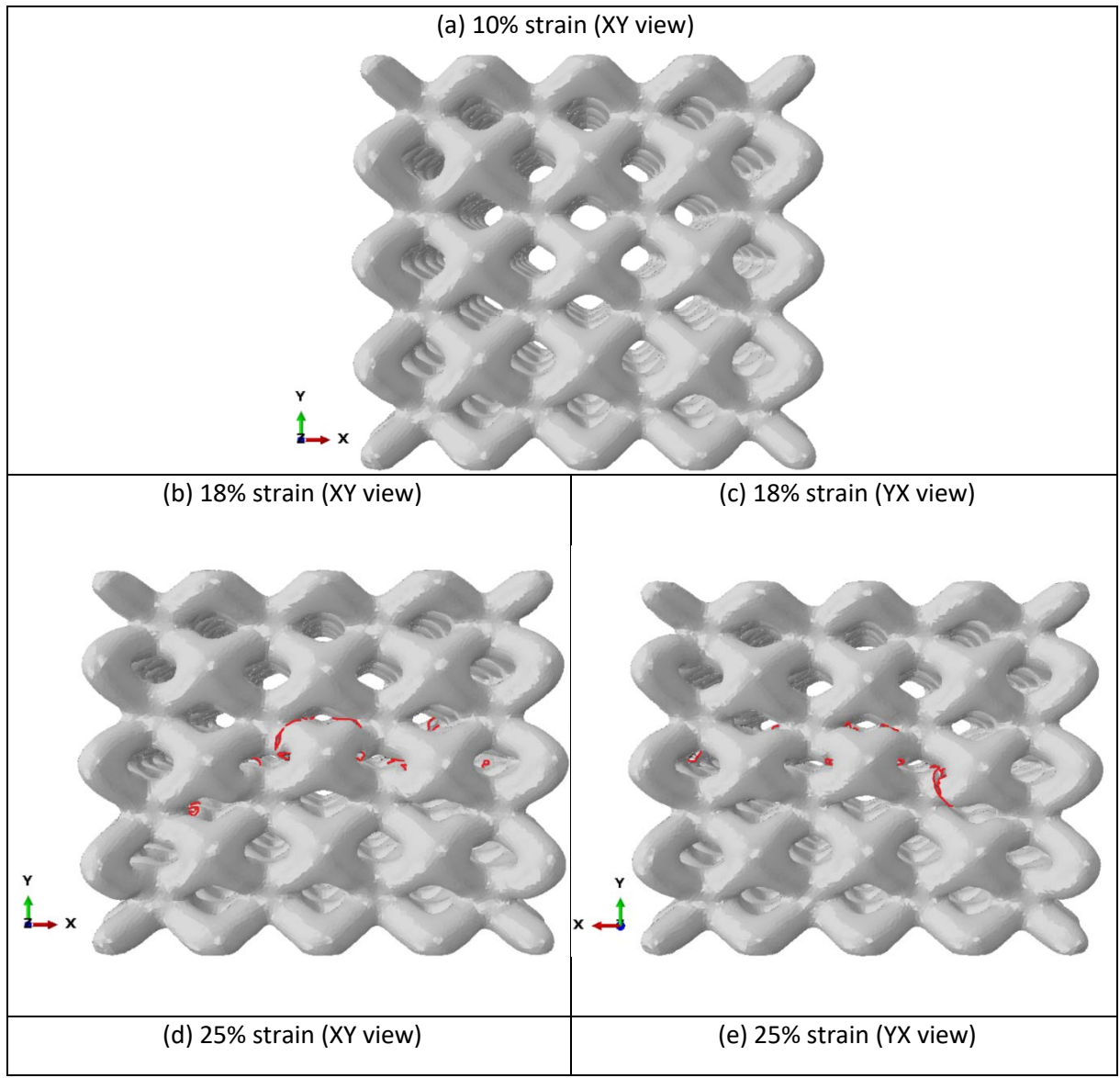

(a) 10% strain (XY view)

(b) 18% strain (XY view)

(c) 18% strain (YX view)

(d) 25% strain (XY view)

(e) 25% strain (YX view)

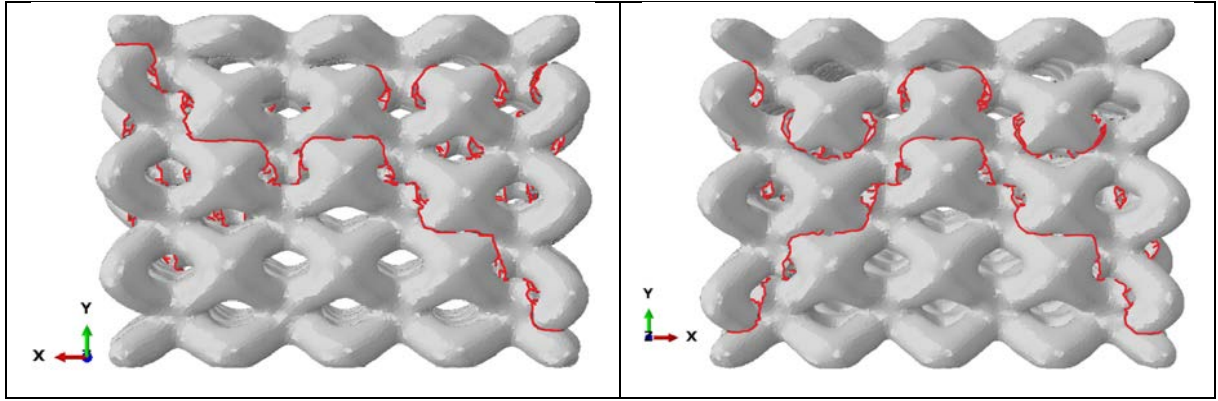

Figure 9. Lateral views of the simulation of the scaffold with nominal strut diameter of 750 μm up to different strains. (a) 10% (close to the minimum stress after the first load drop). (b) and (c) 18% (close to the minimum corresponding to the second load drop). (d) and (e) 25% (close to the minimum corresponding to the second load drop). The cracks in the scaffold have been highlighted in red.

The serrations in the simulated $S - \varepsilon$ curve are more marked than in the experimental ones due to the perfect symmetry of the model, which favours the layer-by-layer failure process. Very interestingly, the model captures very well the successive maximum and minimum peaks in the stress-strain curve during the compression of the scaffold. A deformation of ~4% separates each consecutive minimum and maximum in the $S-\varepsilon$ curves for both strut diameters that corresponds to the deformation necessary to compact an intermediate layer of the scaffold. Nevertheless, failure in the experiments starts at the top and bottom rows of cells while the central row fails first in the simulations and these differences are associated with the boundary conditions. They are never perfect in the experiments and lead to stress concentrations at the contact between the loading plates and the scaffold. The failure mechanisms in the experiments and simulations of the scaffold with 500 μm strut diameter are equivalent and the corresponding movies are not included for the sake of brevity.

The $S - \varepsilon$ curves obtained from the simulations of the scaffolds with struts of 500 μm after 3, 5 and 7 days of immersion in SBF are plotted in Figs. 10a, 10b and 10c, respectively, together with the experimental ones [12]. The experimental curves show a marked reduction in the strength of the scaffolds with respect to the as-printed one (Fig. 7b), particularly after 7 days of immersion in SBF. Moreover, the serrations in the curves also tend to disappear with the immersion time because localized corrosion destroys the symmetry of the scaffold and the layer-by-layer failure mechanism is no longer dominant.

Three different simulations were carried out for each corroded scaffold because the corrosion rate at each node of the scaffold surface is assigned randomly at the beginning of the simulation. The corresponding $S - \varepsilon$ curves obtained from the numerical simulations are also plotted in Fig. 10 and compared with the experimental ones. Because the reduced number of available scaffolds, mechanical tests were only carried out in one scaffold after 3, 5 and 7 days of immersion in SBF and, therefore, there is no information about the experimental scatter associated with the large variability in the corrosion pattern for same nominal experimental conditions. This limitation may be partially responsible for the differences between experiments and simulations. Overall, the simulations capture the degradation of the mechanical properties of the scaffolds due to corrosion, particularly after 3 and 5 days of immersion in SBF. The experimental elastic modulus in both cases is slightly lower than the one obtained in the simulations for the reasons indicate above but the predictions of the initial yield strength were in good agreement with the experimental values. The simulations predicted very accurately the reduction in the peak stress of the scaffold after 3 days of immersion in SBF but the experimental curve showed a progressive reduction in the load carried by the scaffold afterwards (including several oscillations due to progressive compaction). The simulation showed, however, progressive hardening after the yield strength. The lack of hardening in the experimental curve of the scaffold after 3 days of immersion in SBF could be due to the fall off of struts due to corrosion, which delays the compaction process. On the contrary, the simulations underestimated slightly the experimental peak stress of the scaffold after 5 days in SBF but captured the progressive hardening and the oscillations in the $S - \varepsilon$ curve after the yield strength

The quantitative agreement between experiments and simulations is worse in the case of the scaffolds immersed in SBF during 7 days and the yield strength predicted from the simulations overestimated the experimental one by a factor close to 2. Nevertheless, this large difference may be associated with the fact that there was only one experiment in this condition and the experimental scatter is known to increase with the degradation time. Moreover, the simulations capture the disappearance of the serrations in the $S - \varepsilon$ curves in the scaffolds after 7 days of immersion.

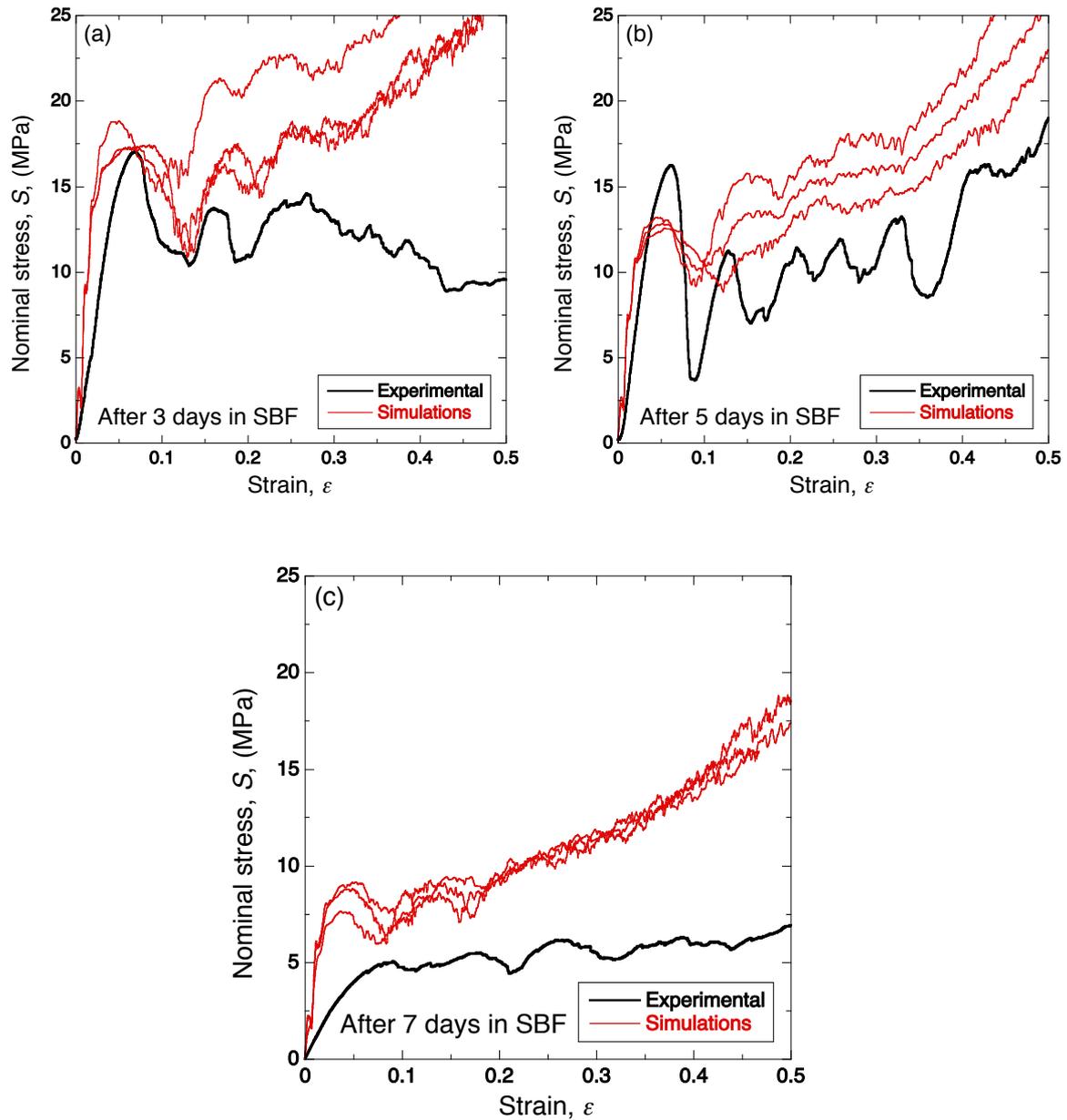

Figure 10. Experimental and simulated nominal stress (*S*) vs. strain (*ε*) curves of the corroded scaffolds with strut diameter of 500 μm deformed in compression. (a) After 3 days of corrosion in SBF. (b) *Idem* 5 days. 8c) *Idem* 7 days.

The deformation and fracture mechanisms of the corroded scaffolds according to the simulations are plotted in Figs. 11 and 12 after 3 and 7 days of immersion in SBF, respectively. The lateral view of the simulated scaffold after 3 days of immersion in SBF is shown in Fig. 10a, where surface damage due to corrosion is easily appreciated. Deformation was homogeneous up to an applied strain of 10%, where the first crack (marked in red in Fig. 11b) appeared at a joint with a large corrosion pit. Further deformation led to the propagation of damage to form a shear band that was oriented at approximately 45º from the loading axis (Fig. 11c). Thus, failure did not happen by a progressive layer-by-layer collapse and the $S - \varepsilon$ curves did not present the large oscillations in stress found in the as-printed scaffolds. After the formation of the shear band, the stress carried by the scaffold in the simulations increased with strain due to the friction and -at longer strains- to the densification of the scaffold. This phenomenon was not observed in the experimental scaffold tested after three days of immersion in SBF (although it was found in the scaffold tested after 5 days of immersion is SBF, Fig. 10b). This difference between experiments and simulations can be attributed to that the experimental scaffold can fall apart after cracking began and, thus, densification was never observed in the $S - \varepsilon$ curve in Fig. 9a.

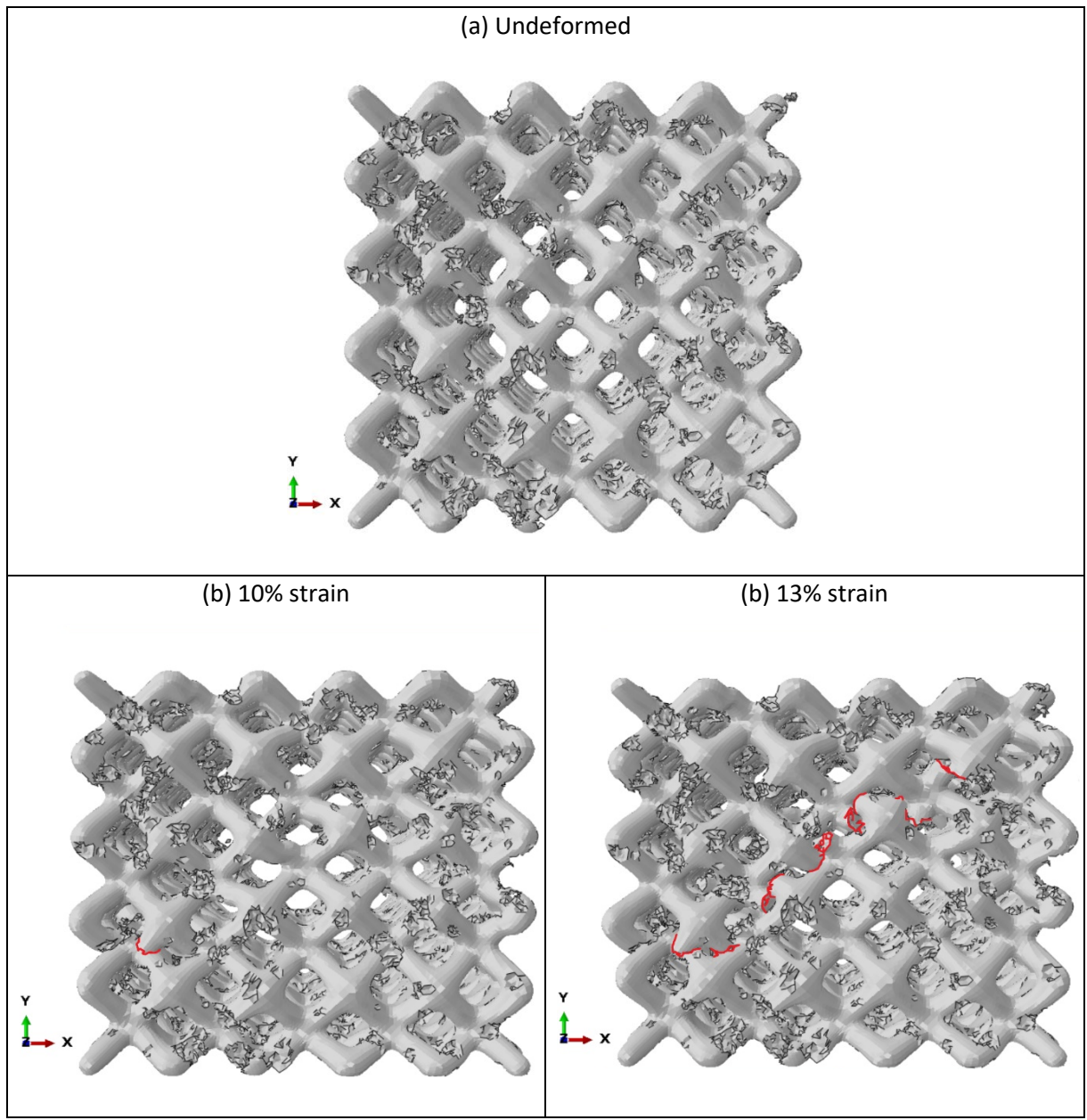

Figure 11. Lateral views of the simulation of the scaffold with nominal strut diameter of 500 μm after 3 days of immersion in SBF up to different strains. (a) Undeformed scaffold with partially degraded struts. (b) Onset of cracking at 10% strain. (c) Formation of a shear band at 45° shear band at 13% strain. The cracks in the scaffold have been highlighted in red.

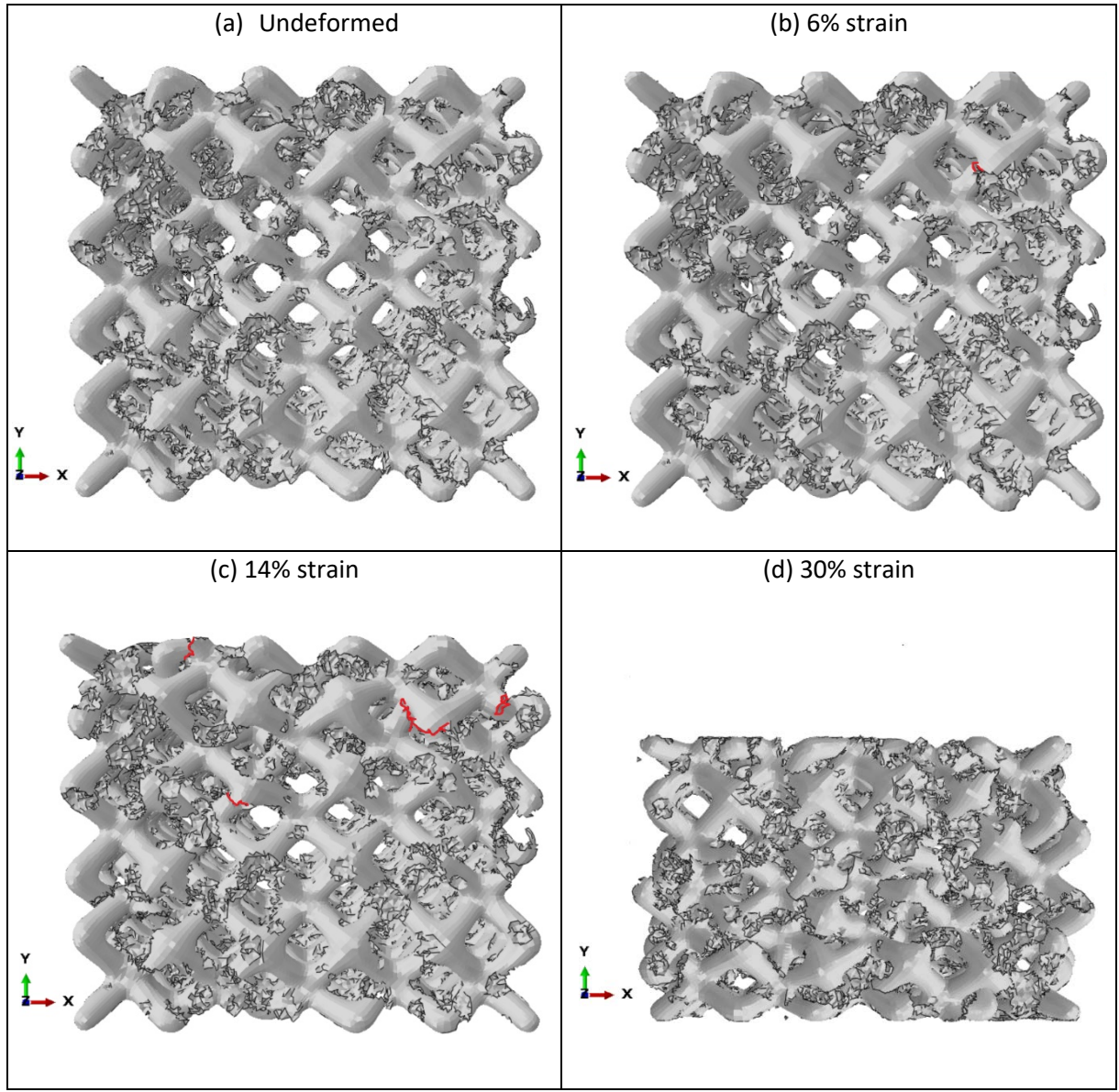

Figure 12. Lateral views of the simulation of the scaffold with nominal strut diameter of 500 μm after 3 days of immersion in SBF up to different strains. (a) Undeformed scaffold with partially degraded struts. (b) Onset of cracking at 6% strain (marked by a yellow arrow). (c) Progressive cracking at 14% strain. (d) Progressive collapse of cells at 30 % strain leading to the beginning of densification. The cracks in the scaffold have been highlighted in red in (b) and (c) but not in (d) because they are everywhere at this point.

The fracture mechanisms in the simulated scaffold after 7 days of immersion in SBF are plotted in Fig. 12. Large corrosion pits are observed in the scaffold surface (Fig. 12a) and the onset of failure occurs at very low strains (6%) by the apparition of a crack near a large corrosion pit close to a joint (Fig. 12b). Further deformation led to the progressive development of cracks in different regions of the scaffolds with very little increment of the load until densification of the fully cracked scaffold started when the applied strain was around 30 %. Thus, failure by the formation of a shear band did not occur in the grossly degraded scaffold and it was replaced by the progressive collapse associated with the formation of many cracks throughout the scaffold.

## 6. Concluding remarks

A simulation strategy was developed to model the corrosion and mechanical properties of biodegradable Mg scaffolds manufactured by laser power bed fusion after immersion in simulated body fluid. The STP file used as input to manufacture the scaffolds was also used to generate the finite element model of the scaffold. Corrosion was simulated through a phenomenological, diffusion-based model and the diffusion process was analyzed using the finite element method. The diffusion flux at each node of the scaffold surface was randomly assigned following a Weibull distribution that depends on two parameters: $\beta$ determines the average corrosion rate while $\gamma$ controls the degree of localization of corrosion, i.e. pitting corrosion. These parameters were calibrated from X-ray computed tomography results obtained WE43 Mg alloy scaffolds which were immersed in simulated body fluid during different times. The phenomenological diffusion model was able to simulate progression of corrosion in the scaffold as well as the degree of localization.

Finite element models for the simulation of the mechanical properties in compression were obtained directly from the corrosion simulations assuming that the elements in which the Mg concentration was equal to or lower than 40% the initial concentration were completely transformed into $Mg(OH)_2$ and have negligible mechanical properties. Thus, they were erased from the model. Mg was assumed to behave as an isotropic, elastic-perfectly plastic solid following the $J_2$ theory of plasticity and fracture was introduced in the simulations through a ductile failure model. The elasto-plastic properties of the WE43Mg alloy as well as the critical strain that dictates the onset of damage were obtained from experimental results in the literature for this alloy while the damage progression parameter was adjusted by comparison of the simulated and experimental compression stress-strain curves of two as-printed scaffolds with different strut dimensions.

The simulations of the as-printed scaffolds were able to reproduce accurately the yield strength of the scaffolds, the development of cracks at the joints between struts subjected to bending during compression and the serrated shape of the stress-strain curves due to the progressive collapse of successive rows of cells. Moreover, the numerical simulations of the corroded scaffolds were able to capture the reduction in the scaffold strength due to corrosion and the progressive change in the deformation mechanisms with the immersion time. After 3 days of immersion in SBF, the scaffolds failed by the formation of a shear band at 45º while collapse was associated with widespread cracking throughout the scaffold after 7 days of immersion in SBF. Thus, the simulations strategy in this paper is able to take into account the reduction in mechanical properties of biodegradable scaffolds as a result of progressive corrosion. This information is important to tailor the scaffolds properties over time.

**Acknowledgements**

This investigation was supported by the European Union's Horizon 2020 research and innovation programme under the Marie Skłodowska-Curie grant agreement No 813869 and from the Madrid regional government, under programme S2018/NMT-4381-MAT4.0-CM. Additional support from the Spanish Ministry of Science and Innovation under program PID2019-109962RB-I00 is gratefully acknowledged.